\documentclass[11pt]{article}
\usepackage{aas_macros,amsmath,amssymb,color,graphics,epsfig}
\usepackage[numbers,square,comma,sort&compress,merge]{natbib}

\usepackage{amsfonts}
\usepackage{float}

\newcommand{\be}{\begin{equation}}
\newcommand{\ee}{\end{equation}}
\newcommand{\bea}{\setlength\arraycolsep{2pt} \begin{eqnarray}}
\newcommand{\eea}{\end{eqnarray}}
\newcommand{\nn}{\nonumber}

\newcommand{\mm}{\mathrm}

\def\ft#1#2{{\textstyle{\frac{\scriptstyle #1}{\scriptstyle #2} } }}
\def\fft#1#2{{\frac{#1}{#2}}}

\def\0{{\sst{(0)}}}
\def\1{{\sst{(1)}}}
\def\2{{\sst{(2)}}}
\def\3{{\sst{(3)}}}
\def\4{{\sst{(4)}}}
\def\5{{\sst{(5)}}}
\def\6{{\sst{(6)}}}
\def\7{{\sst{(7)}}}
\def\8{{\sst{(8)}}}
\def\sst#1{{\scriptscriptstyle #1}}

\begin{document}

\begin{flushright}
\end{flushright}

\vspace{25pt}
\begin{center}
{\large {\bf Emergent Ising symmetry and supercritical fluids}}

\vspace{10pt}
 Hong-Ming Cui$^{1\dagger}$ and Zhong-Ying Fan$^{1\dagger}$

\vspace{10pt}
$^{1\dagger}${ Department of Astrophysics, School of Physics and Material Science, \\
 Guangzhou University, No. 230 Wai Huan Xi Rd, Guangzhou 510006, P.R. China }\\

\vspace{40pt}

\underline{ABSTRACT}
\end{center}
The symmetry of Ising model questions any single crossover scenario for supercritical fluids.  In this work, we firstly study  a pair of thermodynamic crossovers $L^\pm$ analytically for the Van der Waals class fluids. We uncover an emergent $Z_2$ symmetry in addition to the universal scalings in the scaling regime for this class fluids. By using the self-reciprocal property between coexistenct phases, we further establish that under suitable conditions, the Ising symmetry generally emerges in the scaling regime for a general universality class. As a consequence, the thermodynamic crossovers $L^\pm$ generally exhibit an emergent $Z_2$ symmetry in the scaling regime. This partly resolves the symmetry puzzle raised by the Ising model. The results also imply that the physical importance of the Ising model in critical phenomenon is far beyond the scope of magentic transitions. 


\vfill {\footnotesize  Email: fanzhy@gzhu.edu.cn\,.}

\thispagestyle{empty}

\pagebreak

\tableofcontents
\addtocontents{toc}{\protect\setcounter{tocdepth}{2}}



\section{Introduction}
It is known that the liquid-gas transition does not occur beyond the critical temperature and hence no liquid states or gas states can exist independently for a supercritical fluid. However, to interpret the transition below the critical point appropriately, it is generally proposed that there exists a line of supercritical crossover, which separates the fluids into the liquid-like and the gas-like states, for example the Widom line \cite{xu2005relation, simeoni2010widom, ruppeiner2012thermodynamic, luo2014behavior, corradini2015widom, gallo2014widom, de2021widom} and the Frenkel line \cite{brazhkin2012two, yoon2018two, bolmatov2014structural, bolmatov2015frenkel, fomin2018dynamics, yang2015frenkel}. However, as pointed out in \cite{Li_2024}, the existence of a single crossover line is inconsistent with the symmetry of the Ising model. In that case, the system exhibits a $Z_2$ symmetry between the positive ferromagnetic-like phase (magnetization $M>0$) and the negative ferromagnetic-like phase (magnetization $M<0$). This implies that there must exist a pair of supercritical crossovers, which are symmetric with respect to the critical isochore ( corresponding to a vanishing external magnetic field $H=0$) and separate the supercritical fluids into three (rather than two) phases: the positive ferromagnetic-like, the negative ferromagnetic-like and the indistinguishable states. By defining the crossover lines $L^\pm$ as the maxima of magnetic susceptibility $\chi=\big(\fft{\partial M}{\partial H} \big)_T$ at constant $H$,  it was shown in \cite{Li_2024} that along the $L^\pm$ lines, the external field and the order parameter obey the critical scalings
\be |H^\pm|\sim (T-T_c)^{\beta+\gamma}\,,\qquad |M^\pm|\sim (T-T_c)^\beta  \,,\ee
where $\beta$ and $\gamma$ are universal critical exponents of the Ising universality class.

As a comparison, the liquid-gas transition is not associated to the change of long range structural order. Nevertheless, the liquid-gas critical point and the magnetic transition in the 3D Ising model belong to the same $O(1)$ universality class and display the same critical scalings \cite{kadanoff1976scaling, fisher1983scaling}. Then it was assumed in \cite{Li_2024} that the existence of a pair of thermodynamic crossovers $L^\pm$ is valid to the liquid-gas systems as well. This can be made more precisely by identifying the pressure difference $P_{ex}\equiv P(v\,,T)-P(v_c\,,T)$ to the external field $H$ and the order parameter $\delta v\equiv v-v_c$ to the magnetization $M$, where $v$ is the specific volume and $v_c$ its critical value. Defining the $L^\pm$ lines  by the maxima of the isothermal compressibility $\kappa_T\equiv -\fft{1}{v}\big(\fft{\partial v}{\partial P}\big)_T$ (or the susceptibility $\beta_T=\big(\fft{\partial v}{\partial P}\big)_T$), it was established \cite{Li_2024} that $P_{ex}$ and $\delta v$ obey the same critical scalings as the Ising model. The crossover lines $L^\pm$ are interpreted to separate the supercritical fluids into the liquid-like, the gas-like and the liquid-gas indistinguishable states.

In a latest work \cite{Wang:2025ctk}, examination of the thermodynamic crossovers $L^\pm$ is extended to the holographic fluids dual to the charged black hole in anti-de Sitter (AdS) spacetimes.  The black hole admits a first order small-large black hole transition below a critical temperature, which displays classical critical behaviors  analogous to the Van der Waals (VdW) fluid \cite{Kubiznak:2012wp}. It was established \cite{Wang:2025ctk} that the crossovers $L^\pm$ exhibits the same critical scalings as the VdW fluids.

However, these works heavily rely on numerical calculations so that some peculiar (and universal) features of the thermodynamic crossovers $L^\pm$ are not discovered unfortunately. In particular, we are aware of that the symmetry puzzle raised by the Ising model essentially remains unresolved. This inspires us to explore the issue further by developing an analytical approach.

In this work, we firstly study the thermodynamic crossovers $L^\pm$ analytically for the VdW class fluids. We uncover an emergent $Z_2$ symmetry exactly in addition to the universal scalings in the scaling regime for this class fluids. Remarkably, by using the self-reciprocal property between coexistent phases, we further establish that under suitable conditions, the Ising symmetry  emerges in the scaling regime for a general universality class. As a consequence, the thermodynamic crossovers $L^\pm$ generally exhibit an emergent $Z_2$ symmetry in the scaling regime. This partly resolves the symmetry puzzle raised by the Ising model. The results also imply that the physical importance of the Ising model in critical phenomenon is far beyond the scope of magnetic transitions.

The remainder of this paper is organized as follows. In section 2, we study the thermodynamic crossovers $L^\pm$ for the VdW class fluids. We analytically derive the crossovers $L^\pm$ for the VdW fluid and the charged AdS black hole. We uncover various novel features from these explicit examples. In particular, we establish that the lines $L^\pm$ exhibit an emergent $Z_2$ symmetry in the scaling regime for this universality class. In section 3, we further explore whether the symmetry result is valid to a general universality class. We establish that under suitable conditions, the Ising symmetry indeed emerges in the scaling regime for a general universality class. As a consequence, the thermodynamic crossovers $L^\pm$ generally exhibit an emergent $Z_2$ symmetry in the scaling regime. We also study some  counter examples, in which the Ising symmetry breaks in the critical domain.
We conclude briefly in section 4.

\section{Thermodynamic crossovers for Van der Waals class }

\subsection{Van der Waals fluid}
Consider the Van der Waals (VdW) fluid at first. The equation of state (Eos) reads  \cite{Kubiznak:2012wp}
\bea
\left(P+\fft{a}{v^2}\right)(v+b)=kT\,,
\eea
where  $P$ is the pressure, $v$ the specific volume, $T$  the temperature, and $k$ the Boltzmann constant.  The constant $b>0$ describes the nonzero size corrections from the molecules whereas the constant $a>0$ is a measure of the attraction between them. The critical point  can be read off from the inflection point condition $\fft{\partial P}{\partial v}=\fft{\partial^2 P}{\partial v^2}=0$ \cite{Kubiznak:2012wp} 
\bea
T_c=\fft{8a}{27bk}\,,\quad v_c=3b\,,\quad P_c=\fft{a}{27b^2}\,.
\eea
For convenience, we work with normalized quantities $\hat{t}=T/T_c\,, z=v/v_c\,,p=P/P_c$, in the remaining of this work. The law of corresponding states (Locs) reads
\bea
8\hat{t}=(3z-1)\left(p+\fft{3}{z^2}\right)\,.
\eea
Below the critical temperature, a first order transition occurs between the liquid-gas phases, with the coexistence line given by \cite{Cui:2025bfr}
\bea
&&\hat{t}_*=\fft{27\Big((x+1)\mm{ln}x-2(x-1)\Big)\big(x^2-2x\,\mm{ln}x-1\big)^2}{8(x-1)\big(x-\mm{ln}x-1\big)^2\big(x\,\mm{ln}x-x+1\big)^2}\,,\nn\\
&&p_*=\fft{27x\Big((x-1)^2-x\,\mm{ln}^2x\Big)\Big((x+1)\mm{ln}x-2(x-1)\Big)^2}{(x-1)^2\big(x-\mm{ln}x-1\big)^2\big(x\,\mm{ln}x-x+1\big)^2}\,,
\eea
where the variable $x$ characterizes the two phases. One has
\be z_*=\fft{(x-1)(x-\mm{ln}x-1)}{3(x+1)\mm{ln}x-6(x-1)} \,.\label{zstarvdw}\ee
It follows that  $z_*\leq 1$ for $x\leq 1$ and $z_*\geq 1$ for $x\geq 1$. Therefore, the coexistence line on the T-P plane  can be read off from either the liquid phase, corresponding to $x\leq 1$ or the gas phase, corresponding to $x\geq 1$.


To study the  thermodynamic crossovers $L^{\pm}$, we follow \cite{Li_2024} and extend the coexistence line to the supercritical regime by using the critical isochore
\bea
p(z_c,\hat{t}\,)=4\hat{t}-3\,.
\eea
Taking this as the base line and considering the pressure on the $L^{\pm}$ lines
\be
p^{\pm}_{ex}\equiv p^{\pm}-p(z_c,\hat{t}\,)\,,
\ee
we would like to derive the temperature $T_{\max}$, at which the isothermal compressibility 
$\kappa_T$ takes the maximum for a given $ p^{\pm}_{ex}$. Notice the choice of a thermodynamic response function is not unique. We will return to this point later.

\begin{figure}
\centering
\includegraphics[width=300pt]{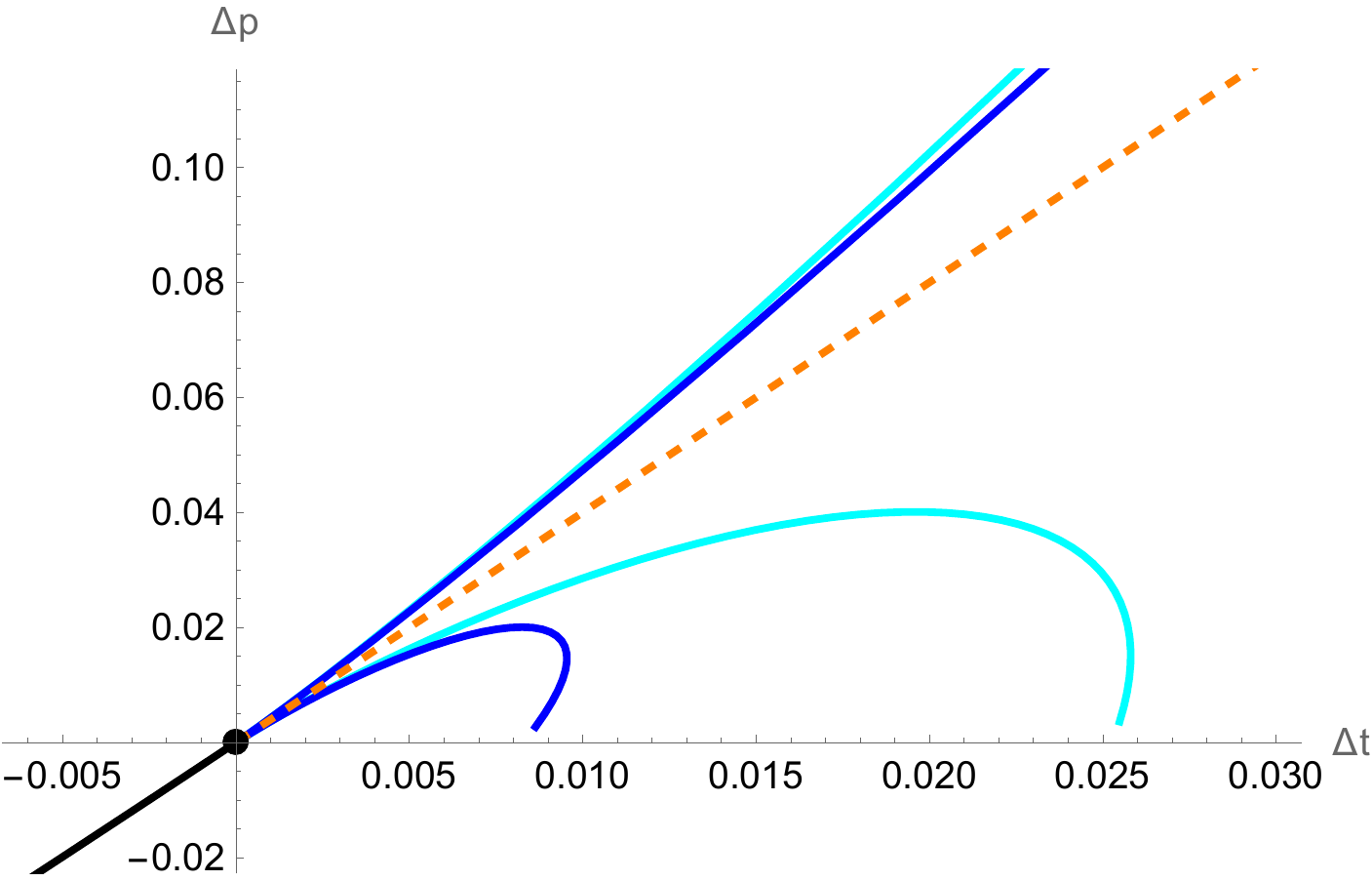}
\caption{Thermodynamic crossovers $L^\pm$ for the VdW fluid $\Delta p=p-1\,,\Delta t=\hat{t}-1$. The black line is the coexistence curve. The cyan (blue) lines are the crossovers defined by using the response function $\kappa_T$ ($\beta_T$).  The dashed line stands for the critical isochore.}
\label{vdw}
\end{figure}

 We firstly compute the isothermal compressibility 
 \bea
\kappa_T=-\fft{1}{z}\left.\fft{\partial z}{\partial p}\right|_T=\fft{z^2(1-3z)^2}{6\left(4tz^3-9z^2+6z-1\right)}\,.
\eea
Notice that for a given pressure difference $p_{ex}$, the normalized specific volume $z$ turns out to be a function of the temperature according to the Locs
\bea\label{eosvdw1}
\hat{t}=-\fft{(3z-1)(3-3z^2+z^2 p_{ex})}{12z^2(z-1)}\,.
\eea
This leads to
\be \left.\fft{\partial z}{\partial \hat{t}}\right|_{p_{ex}}=-\fft{4z^3(z-1)}{(4\hat{t}+p_{ex}-3)z^3-3z+2} \,.\ee
Using this and evaluating the derivative of $\kappa_T$ with respect to the temperature yields
\bea
\left.\fft{d\kappa_T}{d\hat t}\right|_{p_{ex}}=-\fft{2z^4(3z-1)^2\Big(2z^3(3z^2-1) t+(z-1)^2(6z^3-15z^2+7z-1) \Big)}{3\Big( 4z^3(1+t)-9z^2+6z-1 \Big)^3}\,,
\eea
where $t\equiv \hat{t}-1$. It is immediately to read off the  temperature at which $\kappa_T$ peaks as well as the pressure difference $p_{ex}$. One finds
\bea\label{VdW kappa}
&&t=-\fft{(z-1)^2(6z^3-15z^2+7z-1)}{2z^3(3z^2-1)}\,,\nn\\
&&p^{\pm}_{ex}=\fft{3(z-1)^3(3z^2-9z+2)}{z^3(3z^2-1)}\,,
\eea
where “$\pm$” corresponds to $z<1$ and $z>1$ respectively. This specifies a two-branch curve, corresponding to the thermodynamic crossovers $L^{\pm}$ respectively. Their behaviors are depicted in Fig. \ref{vdw}. There are several characteristic features.\\
$\bullet$ In the critical domain, one has to leading order
\bea
p_{ex}=-6\epsilon^3\,,\quad t=\fft{3}{4}\epsilon^2\,,
\eea
where $\epsilon>0$ is associated to the order parameter
\bea
 \omega^{\pm}\equiv z-1=\mp\epsilon\,.
\eea
This implies
\bea\label{VdW scaling}
p^{\pm}_{ex}=\pm \ft{16}{\sqrt{3}}\,t^{3/2}\,, \quad \omega^{\pm}=\mp \ft{2}{\sqrt{3}}\, t^{1/2}\,,
\eea
which is perfectly matched with the critical scalings of $L^{\pm}$ lines for the mean field theory, with exponent $\beta=1/2\,,\gamma=1$ \cite{Li_2024}. Besides, it also produces the correct sign for $p_{ex}$ reported in numerical studies \cite{Li_2024}: $\kappa_T $ peaks at $t_{max}^+$ for a given $p_{ex}>0$ and at $t_{max}^-$ for a given $p_{ex}<0$. Here our new findings is for the VdW fluid, the crossovers $L^\pm$ exhibit an emergent $Z_2$ symmetry (with respect to the critical isochore) in the critical domain. We will show later that this is in fact a universal feature for all fluids in the same universality class.\\
$\bullet$ Away from the critical domain, as temperature increases, the increase of the pressure difference $|p_{ex}|$ on the $L^+$ line is roughly linear and is much slower than that on the $L^-$ line. On the $L^-$ line,  there exists a turning point  beyond which $p_{ex}$ is not unique for a given temperature.  These features support qualitatively the interpretation of the thermodynamic crossover $L^+$ ($L^-$) as the boundary of the liquid-like (the gas-like) states. \\
$\bullet$ Defining the thermodynamic crossovers using the susceptibility $\beta_T=\big(\fft{\partial z}{\partial p}\big)_T$, the behaviors of $L^\pm$ generally change away from the critical domain, see the blue lines in Fig. \ref{vdw}. One has
\bea
&&t=-\fft{(z-1)^2(24 z^3-49 z^2+22 z-3)}{8 z^4(3 z-2)}\,,\nn\\
&&p^{\pm}_{ex}=\fft{3(z-1)^3(6 z^2-13 z+3)}{2 z^4(3 z-2)}\,.
\eea
However, the above features are still valid. In particular, they converge to exactly the same scalings (\ref{VdW scaling}) in the critical domain. On one hand, this tells us that definition of the thermodynamic crossovers $L^\pm$ has something reasonable. In particular, the critical scalings are robust against the choice of a response function. On the other hand, the behaviors of the crossovers $L^\pm$ away from the critical domain imply that their precise and universal definition remains an open question. The situation is quite similar to the single thermodynamic crossover: the Widom line. However, we are not trying to address the issue in this work and leave it to the near future.


\subsection{Charged AdS black hole}
To study whether the above features hold for all VdW-class fluids, we consider holographic fluids dual to the charged AdS black hole in the $D=4$ dimensions \cite{Cai:1998vy, Chamblin:1999tk}.
The equation of state in the extended phase space reads \cite{Kubiznak:2012wp,Gunasekaran:2012dq}
\be P=\fft{T}{v}-\fft{1}{2\pi v^2}+\fft{2Q^2}{\pi v^4} \,,\ee
where $P$ is the thermodynamic pressure defined by  $P=-\Lambda/8\pi$ and $v=2r_h$ stands for the specific volume of black hole molecules. Here $\Lambda$ is the cosmological constant, $r_h$ the horizon radius and $Q$ the electric charge carried by the black hole. It turns out that the small-large black hole transition in the canonical ensemble  in many respects is analogous to that of the VdW fluid. The critical point occurs at \cite{Kubiznak:2012wp,Gunasekaran:2012dq}
\be
v_{c}=2\sqrt{6}\,Q\,,\quad T_c=\fft{\sqrt{6}}{18\pi Q}\,,\quad P_c=\fft{1}{96\pi Q^2}\,.
\ee
The Locs is given by
\be p=\fft{8\hat t}{3z}-\fft{2}{z^2}+\fft{1}{3 z^4}  \,.\ee
Below the critical point, the transition occurs along the coexistence line \cite{Spallucci:2013osa}
\be \hat{t}_*=\sqrt{ \fft{p_*( 3-\sqrt{p_*} )}{2} }  \,.\ee 
To study the thermodynamic crossovers $L^{\pm}$, we firstly write down the critical isochore 
\be p(z_c\,,\hat t\,)=\fft{8\hat{t}-5}{3} \,,\ee
and consider the pressure difference between the $L^\pm$ lines and the critical isochore
\be p^{\pm}_{ex}=p^\pm-p(z_c\,,\hat t\,) \,.\ee
To examine the temperature at which $\kappa_T$ takes the maximum  for a given $p^\pm_{ex}$, we compute the isothermal compressibility 
\be \kappa_T=\fft{3z^4}{8\hat{t}z^3-12 z^2+4} \,,\ee
and  evaluate its derivative with respect to the temperature
\be \fft{d\kappa_T}{d t}\Big|_{p_{ex}}=\fft{-3z^7\Big( 2( t+1)z^4-6z^3+3z^2+4z-3 \Big)}{2\Big( 2(t+1)z^3-3z^2+1 \Big)^3} \,,\ee
where we have adopted the functional relation $z(t)$ for a given $p^\pm_{ex}$, determined by the Locs
\be \big(8\hat t+3p_{ex}-5 \big)z^4-8\hat{t}z^3+6z^2-1=0 \,.\label{eosrn2}\ee
\begin{figure}
\centering
\includegraphics[width=270pt]{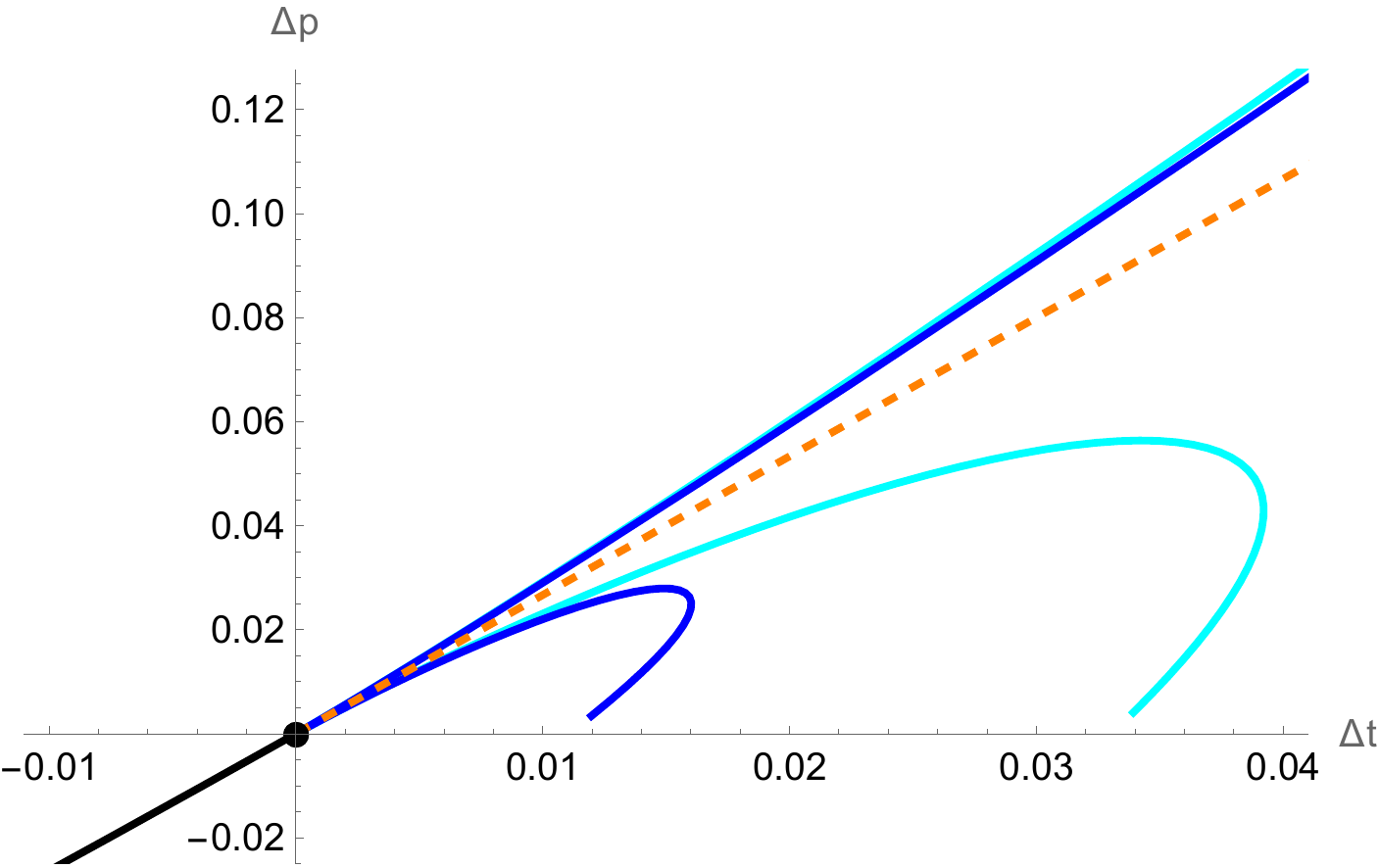}
\caption{The thermodynamic crossovers  for the charged AdS black hole in the canonical ensemble $\Delta p=p-1\,,\Delta t=\hat t-1$. The black line is the coexistence curve. The cyan (blue) lines are the supercritical crossovers defined by using the response function $\kappa_T$ ($\beta_T$).  The dashed line stands for the critical isochore.}
\label{RN}\end{figure}
It is straightforward to read off the temperatures at which $\kappa_T$ peaks as well as the pressure difference \bea\label{crossoverkappa}
&& t=-\fft{(z-1)^2(2z^2-2z-3)}{2z^4}\,,\nn\\
&& p^\pm_{ex}=\fft{(z-1)^3(5z^2-9z-12)}{3z^5}\,,
\eea
where again ``$\pm$'' corresponds to $z<1$ and $z>1$, respectively. This gives the thermodynamic crossovers $L^{\pm}$ for the supercritical charged AdS black hole. Their behaviors are depicted in Fig. \ref{RN}. Interestingly, the various features of the $L^\pm$ lines are analogous to those of the VdW fluid.  In particular, in the critical domain
\be  p^{\pm}_{ex}= \pm\fft{16}{3}\epsilon^3\,,\quad t = \fft{3}{2}\epsilon^{2} \,,\ee
where $\epsilon>0$ and $ \omega^{\pm}\equiv z-1=\mp\epsilon $.  
This implies 
\be  p_{ex}^\pm=\pm \ft{32\sqrt{6}}{27}\, t^{3/2}\,,\qquad  \omega^{\pm}=\mp\ft{\sqrt{6}}{3}\, t^{1/2} \,.\label{RNscale}\ee
This is perfectly matched with the critical scalings of $L^\pm$ lines for the charged AdS black hole, which was firstly studied numerically in \cite{Wang:2025ctk} . Here our new contribution is the $L^\pm$ lines obey an emergent $Z_2$ symmetry in the critical domain, as the VdW fluid.

 Again defining the thermodynamic crossovers using the susceptibility $\beta_T$, the behaviors of $L^\pm$ lines generally change away from the critical point. 
 One has
\bea
&&  t=-\fft{(z-1)^2(4z^2-3z-4)}{2z^3(2z-1)}\,,\nn\\
&& p_{ex}^\pm=\fft{(z-1)^3(10z^2-11z-15)}{3z^4(2z-1)}\,.
\eea
 However, the various characteristic features are still valid, including exactly the same scalings in the critical domain.

It is easy to see that our derivations for the thermodynamic crossovers $L^\pm$ can be straightforwardly generalized to holographic fluids dual to general AdS black holes, as long as the equation of states are known analytically. We present some more examples in the appendix. However, rather than studying it in a case-by-case basis, we would like to explore the  universal features of the $L^\pm$ lines in the critical domain for all VdW class fluids.

\subsection{Universal analysis in the critical domain}

In the critical domain, the equation of states share universal features for fluids in the same universality class, known as the scaling form, which relies on various critical exponents $\{\beta\,,\gamma\,,\delta\}$. Hence we are looking forward to uncovering universal features in the critical domain for the thermodynamic crossovers $L^\pm$. 

The VdW class fluids have the exponents $\beta=1/2\,,\gamma=1\,,\delta=3$. Close to the critical point, the law of corresponding states can be expanded as \cite{Hu:2024ldp}
\be p(\omega\,,t)=1+\sum_{i\geq 1}a_{i0}\, t^i+\sum_{j\geq 3}a_{0j} \omega^j+\sum_{i\,,j\geq 1}a_{ij}\, t^i\,  \omega^j \,.\label{geneeos}\ee
 Notice that $a_{0j}=0$ for $j\leq 2$ and $a_{11}\neq 0$ essentially guarantees the values of the various exponents. It should not be confused that the above expansion is valid to both the supercritical and the coexistent regions but the series coefficients generally might be different in the two cases. Here for our purpose, we focus on the supercritical region.

From the expansion, the critical isochore  is specified by
\be p(\omega_c\,,t)=1+\sum_{i\geq 0}a_{i0}\, t^i \,.\ee
Hence we can write
\bea \label{genedeltap}
p_{ex}&=&p(\omega\,,t)-p(\omega_c\,,t)    \nn\\
&=&a_{11}\, t\, \omega+a_{12}\, t\, \omega^2+a_{21}\, t^2\, \omega+a_{03}\, \omega^3+\cdots\,,
\eea
where we have dropped the higher order terms, which are irrelevant to our discussions. To derive the thermodynamic crossovers $L^\pm$, we need extract the temperature at which the response function $\kappa_T$ peaks for a fixing $p_{ex}$. The first step toward this is to find 
the functional relation $\omega(t)$ when $p_{ex}$ is held fixed. According to (\ref{genedeltap}), we deduce
\be \left.\fft{d \omega}{d t}\right|_{ p_{ex} }=-\fft{\omega\big(a_{11}+a_{12} \omega+2a_{21} t\big) }{\big( a_{11} t+2a_{12} t \omega+a_{21} t^2+3a_{03} \omega^2 \big)} \,.\label{omegat}\ee
  To proceed, we compute the isothermal compressibility $\kappa_T$ by using (\ref{geneeos})
\be dp=3a_{03}\, \omega^2 d \omega\qquad \Rightarrow\qquad \kappa_T=-\fft{1}{a_{11} t+2a_{12} t \omega+a_{21} t^2+3a_{03} \omega^2 }\,.\ee
Then using (\ref{omegat})  and differentiating $\kappa_T$ with respect to the temperature yields
\be \left.\fft{d \kappa_T}{d  t}\right|_{ p_{ex} }=\kappa_T^3\Big(a_{11}^2 t-3a_{03} a_{11} \omega^2+\cdots \Big) \,.\ee  
Clearly in the critical domain the temperature at which the isothermal compressibility peaks is given by
\be  t=\fft{3a_{03} \omega^2}{a_{11}} \,. \ee
By plugging this into (\ref{genedeltap}), we arrive at 
\be  p_{ex} =4a_{03} \omega^3 \,.\ee
These results imply on the thermodynamic crossovers $L^\pm$
\be p_{ex}^\pm \Big|_{L^\pm }=\pm c_1\, t^{3/2}\,,\qquad \omega^\pm\Big|_{L^\pm }=\mp\,c_2\, t^{1/2} \,,\label{deltapuniv}\ee
where $c_1\,,c_2$ are two scale factors that are material dependent and differentiate for fluids in the same universality class. One has
\be c_1=-4a_{03}c_2^3\,,\qquad c_2=\sqrt{\fft{a_{11}}{3a_{03}}} \,.\label{c12vd2}\ee
The results can be tested for a number of examples, including the VdW fluid and holographic fluids dual to various AdS black holes examined in this work.\\
Several comments are in order:\\
$\bullet$ Our results not only predict universal scalings for the thermodynamic crossovers $L^\pm$ for the VdW class fluids but also establish an emergent Z$_2$ symmetry in the critical domain. This is surprising since it is generally considered to a characteristic feature of the Ising model. This inspires us to study the general situations, see the next section.   \\
$\bullet$ In above derivations, the $a_{12}$ and $a_{21}$ terms are subleadings and hence are unimportant because of $ \omega^\pm\propto  t^{1/2}$. The situation is exactly the same as the derivations of critical exponents in the coexistence region \cite{Kubiznak:2012wp}. In the latter case, these terms are actually outside the scaling regime. In viewing of this, we expect that the critical scalings of the thermodynamic crossovers $L^\pm$ depends only on the scaling form of the equation of states. This is very important when studying a general universality class.\\
$\bullet$ While the thermodynamic crossovers $L^\pm$ could be defined by the susceptibility $\beta_T$, one has $\beta_T=-\kappa_T$ to leading order in the critical domain. Hence it will inevitably give rise to the same results in the scaling regime.

\section{Emergent Ising symmetry in critical phenomenon}
The above results for VdW class fluids inspire us to study whether the thermodynamic crossovers $L^\pm$ exhibit an emergent $Z_2$ symmetry for a general universality class. In this section, we will show that under suitable conditions, the Ising symmetry emerges in critical phenomenon for a general quantum fluid. This will partly resolve the symmetry puzzle raised by the Ising model. As a consequence, the thermodynamic crossovers $L^\pm$ generally exhibit a $Z_2$ symmetry in the critical domain.

\subsection{Emergent Ising symmetry }
It turns out that for a general quantum fluid, there exists a hidden symmetry between the two coexistent phases \cite{Cui:2025bfr}. Without loss of generality, consider the liquid-gas transition. The functional relation between the specific volumes of the liquid-gas phases obeys
\be z_l=\varphi(z_g)\,,\qquad z_g=\varphi(z_l) \,,\label{zlg}\ee
where the subscripts $l\,,g$ denotes the liquid and the gas phases respectively and the function $\varphi$ is called self-reciprocal because of $\varphi=\varphi^{-1}$ \cite{Cui:2025bfr}. This nice property enables one to solve the coexistence line analytically in general situations \cite{Cui:2025bfr}.

Now we apply it to the critical domain. Assume the function $\varphi$ is of $\mathcal{C}^1$ at least. By expanding the relations (\ref{zlg}) near the critical point, one finds to leading order the order parameter $\omega=z-1$ obeys 
\be \omega_l=\varphi'(1)\omega_g\,,\qquad \omega_g=\varphi'(1)\omega_l  \,,\ee
which implies $\varphi'(1)=\pm 1$. Clearly the nontrivial solution is $\varphi'(1)=-1$. Therefore, in the scaling regime, the order parameter obeys
\be \omega_{g\,,l}=\pm A\, t^\beta \,,\label{orderz2}\ee
where $A$ is a underdetermined  constant. It should not be confused that in this subsection $t\equiv 1-T/T_c$. 

To proceed, consider the equation of states in the critical domain, which relies on the various critical exponents $\{\alpha\,,\beta\,,\gamma\,,\delta\}$.
Constrained by the scaling laws, only two of the exponents are truly independent.  Without loss of generality, we choose $\beta\,,\delta$ to be the independent ones. We may write
\be p=p(\omega_c\,,t)+a\,\omega^\delta+\sum_{\delta>i \geq 1}b_i\, t^{\beta(\delta-i)}\,\omega^i+\cdots \,,\ee
where $p(\omega_c\,,t)=1+\sum_i a_i\,  t^i$ stands for the critical isochore. Notice that each term associated to the series coefficient $b_i$'s  will be in the same order of $t$ in the scaling regime because of $\omega\propto t^\beta$. In fact, the behavior of the pressure follows from the scaling form
\be p_{ex}=\omega^\delta\psi(\mu) \,,\quad \mu=t^\beta\omega^{-1} \,,\ee
where $p_{ex}=p-p(\omega_c\,,t)$ defines the external field.  We demand the equation of states is regular at $t=0$ for $\omega\neq 0$ and $\omega=0$ for $t>0$. This implies that the scaling function $\psi(\mu)$ should be truncated at the order $\mu^\delta$ in the critical domain (this picks up the relevant terms in the scaling regime). In addition, the stability condition for a thermodynamically system  requires 
\be \fft{\partial^3p}{\partial\omega^3}\Big|_{t=0}=a \delta(\delta-1)(\delta-2)\omega^{\delta-3}<0 \,.\label{stability}\ee 
 It is immediately seen that the exponent $\delta$ should be an odd integer 
\be \delta=2N+1\,,\quad N\geq 1 \,,\label{deltaregion}\ee 
and $\psi(0)=a<0$. Otherwise, for an even $\delta$, the coefficient $a<0$ for the gas phase and $a>0$ for the liquid phase, breaking the regularity condition. 

Blow the critical point, the transition occurs at an isobar for a given temperature $p(\omega_g\,,t)=p(\omega_l\,,t) \,,$ leading to
\be
a\,\omega_g^{\delta}+\sum_{\delta>i\geq 1}b_i\, t^{\beta(\delta-i)}\omega_g^i=a\,\omega_l^{\delta}+\sum_{\delta>j\geq 1}b_j\, t^{\beta(\delta-j)}\omega_l^j\,,\label{isobar} \ee
and obeys the Maxwell's area law $\int_{\omega_l}^{\omega_g}\omega dp=0$, which gives
\be
\fft{\delta\,a}{\delta+1}\,\omega_g^{\delta+1}+\sum_{\delta>i\geq 1}\fft{ib_i}{i+1}\, t^{\beta(\delta-i)}\omega_g^{i+1}= \fft{\delta\,a}{\delta+1}\,\omega_l^{\delta+1}+\sum_{\delta>j\geq 1}\fft{jb_j}{j+1}\, t^{\beta(\delta-j)}\omega_l^{j+1} \,.\label{arealaw}
\ee
However, the $Z_2$ symmetry between the order parameter (\ref{orderz2}) implies that only one of the two equations is independent. Then the existence of consistent solutions to the order parameters implies that 
only odd powers of $\omega$ can survive in the scaling regime, namely
\be b_{2k}=0\,,\quad k\geq 1 \,.\ee
In this case, the Maxwell's area law is automatically satisfied and the solution to the order parameter (\ref{orderz2}) will be determined by
\be a\,A^\delta+\sum_{N\geq k\geq 1}b_{2k-1}A^{2k-1}=0 \,.\label{oddA}\ee
It should be emphasized that this does not only give a valid solution to $A$ but also generally puts more constraints on the series coefficients. 


Combing these results, we arrive at that the external field $p_{ex}$ should be an odd function of $\omega$ in the scaling regime
\be  p_{ex}=\sum_{N\geq i\geq 0} b_{2i+1}\,t^{2\beta(N-i)}\,\omega^{2i+1} +\cdots \,,\ee
where dots stands for higher order terms which lose control under the stability condition and hence may have even powers of $\omega$. This generalises the Laudau theory, which is initially developed for the mean field theory. As a consequence, the scaling function should be even $\psi(-\mu)=\psi(\mu)$ and hence is negative definite in the scaling regime. 

We conclude that under suitable conditions, the Ising symmetry emerges for a general quantum fluid in the scaling regime
\be p_{ex}^{\pm}=\pm c_1\,t^{\beta\delta}\,,\qquad  \omega^{\pm}=\mp c_2 \,t^\beta \,,\label{z2coexistence}\ee
where $c_1\,,c_2$ are two scale factors, which are positive and material dependent. It should not be confused that the superscript ``$\pm$'' stands for the two coexistent phases. Notice that in our convention, the product $p_{ex}^\pm\omega^\pm$ is always negative definite, exactly the same as the Ising model. The result (\ref{z2coexistence}) is not only formal but also has its physical origin: the spontaneous symmetry breaking in second order transition.

To see this, consider the Ising model at first. The theory is $Z_2$ symmetric (for a vanishing external magnetic field $H=0$) and the system is in the paramagnetic phase above the critical temperature. Below the critical point, the Ising symmetry is spontaneously broken and the system will be in either the positive-ferromagnetic phase (magnetization $M>0$) or the negative-ferromagnetic phase ($M<0$). The both are thermodynamically preferred on an equal footing and hence can coexist. The magnetization $M$ for the two phases exhibits a $Z_2$ symmetry: $M\rightarrow -M$. This gives a typical example of the self-reciprocal property, corresponding to $\varphi(M)=-M$. In the presence of a nonvanishing external field, the equation of state obeys $H(-M)=-H(M)$. Hence, the Taylor expansion around the critical point has only odd powers of $M$.

As a comparison, the liquid-gas transition is not associated to the change of long range structural order. Despite the difference, the transition is of second order at the critical point and hence  a certain continuous symmetry of the supercritical fluid is spontaneously broken. As a remanent, the self-reciprocal property emerges between the two ordered phases. The story is much like the Ising model, except that the symmetry of the microscopic theory specified by the self-reciprocal function $\varphi(\omega)$ is generally more complicated.  However, the physical picture is unified in the spontaneous symmetry breaking of second order transitions. In particular, in the critical domain $\varphi(\omega)=-\omega$, exactly the same as the Ising model. This suggests that surprisingly, the liquid-gas transtion in the scaling regime can be partly understood from the Ising model by identifying $p_{ex}\rightarrow H\,,\omega\rightarrow M$ macroscopically and $v_i\rightarrow s_i$ microscopically, where $v_i$ is the specific volume of the $i$-th molecule and $s_i$ the magnetic dipole. This is remarkable. Despite that the critical point is singular, the underlying theory for a general quantum fluid is dramatically simplified in the scaling regime. This implies that the physical importance of the Ising model in critical phenomenon is far beyond the scope of magnetic transitions.

\subsection{Thermodynamic crossovers}
Now we return to the thermodynamic crossovers $L^\pm$ defined in the supercritical region. The equation of states can still be written in a scaling form $ p_{ex}=\omega^\delta\tilde{\psi}(\mu)$ in the critical domain \cite{stanley1971}. Here the function $\tilde\psi$ is generally different from $\psi$ but for simplicity, we will omit the tilde in following discussions. Clearly, the emergent Ising symmetry implies that the function $\psi$ is even 
\be \psi(\mu)=\psi(-\mu) \,,\ee
and $\psi(\mu)<0$ in the scaling regime. This property essentially guarantees the emergence of $Z_2$ symmetry for the thermodynamic crossovers $L^\pm$.

Firstly, when $p_{ex}$ is held fixed, the functional relation $\omega(t)$ is determined by
\be \fft{d\omega}{dt}\Big|_{p_{ex}}=-\fft{\omega\beta t^{\beta-1}\psi'(\mu)}{\delta\omega\psi(\mu)-t^{\beta}\psi'(\mu) } \,.\label{scalingomega}\ee
Evaluation of the compressibility yields
\be \kappa_T=-\fft{\partial\omega}{\partial p_{{ex}}}\Big|_t=-\fft{\omega^{2-\delta}}{\delta\omega\psi(\mu)-t^{\beta}\psi'(\mu) } \,.\label{scalingkappa} \ee
Differentiating $\kappa_T$ with respect to $t$ for a fixing $p_{ex}$, we deduce
\be \fft{d\kappa_T}{dt}\Big|_{p_{ex}}=\beta t^{2\beta-1}\omega^{2(\delta-2)}\kappa_T^3\,\Big[\delta\psi(\mu)\psi''(\mu)-(\delta-1)\psi^{'2}(\mu) \Big] \,.\ee
It follows that the temperature at which $\kappa_T$ peaks can be read off from
\be \delta\psi(\mu)\psi''(\mu)-(\delta-1)\psi^{'2}(\mu)=0 \,.\ee
Since $\psi(\mu)$ is even, the $Z_2$ symmetric solution exists 
\be \omega^\pm=\mp c_2\, t^{\beta}\,, \ee
where $c_2>0$, determined by
\be \delta\psi\big(c_2^{-1} \big)\psi''\big(c_2^{-1}\big)-(\delta-1)\psi^{'2}\big(c_2^{-1} \big)=0 \,.\ee
Since the exponent $\delta$ is odd, the external field obeys 
\be p_{ex}^{\pm}=\pm c_1\,t^{\beta\delta}\,, \ee
where 
\be c_1=-c_2^\delta\,\psi\big( c_2^{-1} \big)>0 \,.\label{c1c2gene}\ee
Notice that the positiveness of $c_1$ is guaranteed by the behavior of the function $\psi$ in the scaling regime. This completes our derivations. It is easily seen that the $Z_2$ symmetry between the thermodynamic crossovers $L^\pm$ essentially follows from the emergent symmetry of the system under consideration. The situation is exactly the same as the Ising model. This partly resolves the symmetry puzzle raised by the Ising model. In this sense, the thermodynamic crossovers $L^\pm$ should be distinguished from the Widom line.

Finally, let us give a simplest (but nontrivial) example: $b_1\equiv b\neq 0$ and $b_{2k+1}=0$ for $k\geq 1$. In this case, one has $\psi(\mu)=a+b \mu^{\delta-1}$ and
\be  p_{ex}=a\, \omega^\delta+b \, t^{\beta(\delta-1)}\,\omega+\cdots \,,\label{deltapgene}\ee
where $a<0\,,b<0$. By direct computations, one can reproduce the above results for the crossovers $L^\pm$, with $c_1\,,c_2$ given by
\be c_1=-a(\delta-1)^2\,c_2^\delta\,,\qquad c_2=\left[ \fft{b}{a\delta(\delta-2)} \right]^{1/2N}\,.\ee
This is consistent with (\ref{c1c2gene}). This example includes the VdW class which has $\beta=1/2\,,\delta=3$. In that case, $a=a_{03}\,,b=a_{11}$, one immediately finds that $c_1=-4a_{03}c_2^3\,,c_2=\sqrt{\fft{a_{11}}{3a_{03}}}$, which is well matched with (\ref{c12vd2}).



\subsection{Counter examples}
Having established the emergence of Ising symmetry in critical phenomenon, we shall point out that there exists counter examples: the holographic fluids dual to a class of quantum (corrected) black holes \cite{Hu:2024ldp,Cui:2025qdy}. In that case, one has the exponents $\beta=1\,,\delta=3$. The issue occurs because of an integer $\beta$, which leads to awkward features: the solution to the coexistence line can be analytically continued to $T>T_c$. 

Let us first consider the coexistence region for this universality class. The equation of states close to the critical point can be expanded as 
\be p=p(\omega_c\,,t)+a_{12} t\omega^2+a_{21} t^2\omega+a_{03} \omega^3+\cdots \,,\ee
where $t=1-T/T_c$. Notice that the coefficient $a_{12}$ associated to $\omega^2$ does not vanish because of $\beta=1$. Using the isobar condition $p(\omega_g\,, t)=p(\omega_l\,, t)$ and the Maxwell's area law $\int_{\omega_l}^{ \omega_g}\omega dp=0$, we deduce
\bea
&& \Big[ a_{03}\omega^3+a_{12} t\omega^2+a_{21} t^2\omega \Big]\Big|_{\omega_l}^{\omega_g}=0\,,\nn\\
&&\Big[ \fft34 a_{03}\omega^4+\fft23 a_{12} t\omega^2+\fft12 a_{21} t^2\omega \Big]\Big|_{\omega_l}^{\omega_g}=0 \,.
\eea
It follows that provided 
\be 0<3a_{03}a_{21}<a_{12}^2<\fft92 a_{03}a_{21} \,,\label{counterregion}\ee 
there exists generally non-$Z_2$-symmetric solutions of the form
\be \omega_{g\,,l}=A_{g\,,l}\,t^\beta \,,\label{countercoe}\ee
with 
\be A_{g\,,l}=\fft{-a_{12}\pm\sqrt{3(a_{12}^2-3a_{03}a_{21}  )} }{3a_{03}} \,. \ee
Notice that within the parameter region (\ref{counterregion}), the upper bound on $a_{12}^2$ guarantees $A_g A_l<0$. However,  a nonvanishing $a_{12}$ inevitably leads to $A_g\neq -A_l$, implying that the self-reciprocal function $\varphi$ is not of $\mathcal{C}^1$. As a consequence, on the $T-v$ plane, there will exist a discontinuity in the phase diagram around the critical point \cite{Cui:2025qdy}. This clarifies why the emergent $Z_2$ symmetry breaks. 
However, since $\beta=1$, it turns out that the coexistence solutions (\ref{countercoe}) can be analytically continued to $T>T_c$ by identifying $t\rightarrow -t\,,A_g\rightarrow -A_l\,,A_l\rightarrow -A_g$. This is awkward since it implies that the transition can occur for $T>T_c$ as well.

Next, we move to study the supercritical crossovers $L^\pm$. One has
\be p_{ex}\equiv p-p(\omega_c\,, t)=a_{03}\omega^3+a_{12}t\omega^2+a_{21} t^2\omega +\cdots\,. \ee
Note $t=T/T_c-1$ and we assume the series coefficients are still within the region (\ref{counterregion}). By simple calculations, we obtain
\be \fft{\partial\omega}{\partial t}\Big|_{ p_{ex}}=\big( a_{12}\omega^2+2a_{21} t\omega \big)\kappa_T \,,\ee
 and the compressibility
\be \kappa_T=-\fft{1}{ 3a_{03}\omega^2+2a_{12} t\omega+a_{21} t^2 } \,.\ee 
Evaluation of the derivative of $\kappa_T$ with respect to the temperature yields
\be \fft{\partial\kappa_T}{\partial t}\Big|_{ p_{ex}}=2\kappa_T^3 t\Big[\big(a_{12}^2-3a_{03}a_{21} \big)\omega^2 +a_{12}a_{21} t \omega+a_{21}^2 t^2 \Big] \,.\ee
It follows that within the parameter region  (\ref{counterregion}), real solutions to the lines $L^\pm$ exist
\be  \omega^{\pm}=\fft{ -a_{12}a_{21}\mp \sqrt{3a^2_{21}\big( 4a_{03}a_{21}-a_{12}^2 \big)} } {2\big( a_{12}^2-3a_{03}a_{21} \big)}\, t \,,\ee
and 
\be  p^\pm_{ex}=\fft{ a_{21}^2 \big( 4a_{03}a_{21}-a_{12}^2 \big)} {2\big( a_{12}^2-3a_{03}a_{21} \big)^3}\,\left[ a_{12}\big(2a_{12}^2 -9a_{03}a_{21}\big)\mp 3a_{03}\sqrt{3a^2_{21}\big( 4a_{03}a_{21}-a_{12}^2 \big)}\,\,\right]\, t^3 \,.\ee
However, it turns out that the product of the order parameters and the external fields are always positive definite within the parameter region (\ref{counterregion})
\bea 
&&\omega^+\,\omega^-=\fft{a_{21}^2\,t^2}{a_{12}^2-3a_{03}a_{21}}>0 \,,\nn\\
&&p^+_{ex}\, p^-_{ex}=\fft{\big( a_{12}^2-4a_{03}a_{21}\big)^2 a_{21}^4\,t^6}{\big(a_{12}^2-3a_{03}a_{21} \big)^3}>0 \,.
\eea
This implies that while formal solutions to the lines $L^\pm$ generally exist in the phase space, they lose the meaning of supercritical crossovers, which separate the supercritical fluids into three distinct phases.

Let us return to explicit examples. The first is the quantum anomaly corrected black holes studied in \cite{Hu:2024ldp}. In this case, one has $a_{21}=0$. In this case, while physical solutions exist in the coexistence region (\ref{countercoe}), there does not exist valid solutions to the crossover lines $L^\pm$ any longer. The interpretation of this is simple: the black hole exhibits transitions for both $T<T_c$ and $T>T_c$ and hence no supercritical states exist. The second example is the $U-\nu$ criticality of quantum BTZ black holes studied in \cite{Cui:2025qdy}. In this case $a_{21}=-\ft{1881}{32}\neq 0$ and $a_{03}=-\ft{27}{8}\,,a_{12}=\ft{405}{16}$. The parameters obey the condition (\ref{counterregion}) exactly. However, again the black hole exhibits transitions for both $T<T_c$ and $T>T_c$ and hence no supercritical states exist. 

These examples suggest that there does not exist non-$Z_2$ symmetric supercritical states in the scaling regime. If this is true in general situations, the emergence of Ising symmetry in critical phenomenon is deeply connected to the supercritical states in a way far beyond our expectations. This certainly deserves further investigations.

\section{Conclusions}

The symmetry of Ising model questions any single crossover scenario for supercritical fluids. Recently, Li and Jin introduced a pair of thermodynamic crossovers $L^\pm$, which separate the supercritical fluids into the liquid-like, the gas-like and the indistinguishable states \cite{Li_2024}. However, the symmetry puzzle essentially remains unresolved. 

In this work, we firstly studied the thermodynamic crossovers $L^\pm$ analytically for the VdW class fluids. We uncover an emergent $Z_2$ symmetry exactly in the scaling regime in addition to the universal scalings for this class fluids. This inspires us to explore whether the symmetry emerges for a general universality class. We found that under suitable conditions, the answer is ``yes''.

 The reason is there exists a hidden symmetry between the coexistent phases \cite{Cui:2025bfr}. The specific volumes obey the self-reciprocal property (\ref{zlg}). This enables us to establish that under suitable conditions, the Ising symmetry emerges in the scaling regime for a general universality class. As a consequence, the thermodynamic crossovers $L^\pm$ generally exhibit a $Z_2$ symmetry in the scaling regime. This partly resolves the symmetry puzzle raised by the Ising model. The results also imply that the physical importance of the Ising model in critical phenomenon is far beyond the scope of magnetic transitions. 

 We also studied a class of counter examples: the holographic fluids dual to a class of quantum (corrected) black holes \cite{Hu:2024ldp,Cui:2025qdy}, which has the exponents $\beta=1\,,\delta=3$. The $Z_2$ symmetry breaks because of an integer $\beta$, which leads to awkward features: the solution to the coexistence line exists for $T>T_c$ as well. In other words, no supercritical states exist for this class fluids.  The absence of non-$Z_2$ symmetric $L^\pm$ lines suggests that the emergent Ising symmetry in the critical domain is deeply connected to the supercritical states in a way far beyond our expectations.

\section*{Acknowledgments}

Z.Y. Fan was supported in part by the National Natural Science Foundations of China with Grant No. 11873025.

\appendix

\section{Thermodynamic crossovers for more AdS black holes}

Here we present more results about the thermodynamic crossovers $L^\pm$ for the rotating AdS black hole and the Gauss-Bonnet black hole.

 Consider the rotating AdS black hole at first. We are interested in the leading angular momentum $J$ corrections. The equation of states simplifies to \cite{Gunasekaran:2012dq}
\bea
T=Pv+\fft{1}{2\pi v}-\fft{48J^2}{\pi v^5}+\mathcal{O}(J^4)\,.
\eea
where $v$ is the specific volume of the black hole molecule defined as \cite{Gunasekaran:2012dq}
\be
v\equiv2\left(\fft{2V}{4\pi}\right)^{1/3}\,,
\ee
where $V$ is the thermodynamic volume. In this approximation, the critical point occurs at
\bea
v_c=2\times90^{1/4}\sqrt{J}\,,\quad T_c=\fft{90^{3/4}}{225\pi\sqrt{J}}\,,\quad P_c=\fft{1}{12\sqrt{90}\pi J}\,.
\eea
The Locs is given by
\bea
\hat{t}=\fft{15z^4-1+10p\,z^6}{24 z^5}\,,
\eea
Below the critical point, the coexistence curve can be solved analytically as \cite{Cui:2025bfr}
\bea
&& \hat{t}_*=\fft{5x(x+1)}{2(x^2+3x+1) z_*(x)}\,, \nn\\
&& p_*=\fft{3x^2(3x^2+4x+3)}{2(x^4+4x^3+5x^2+4x+1)z_*^2(x)}\,,
\eea
where the variable $x$ is related to the specific volume as
\be z_*(x)=\fft{(x^4+4x^3+5x^2+4x+1)^{1/4}}{15^{1/4}} \,.\ee
It follows that $z_*\leq 1$ for $x\leq 1$ and $z_*\geq 1$ for $x\geq 1$. Thus, the coexistence line can be read off from either the small black hole phase ($x\leq 1$) or the large black hole phase ($x\geq 1$), respectively.

Derivations of the thermodynamic crossovers $L^{\pm}$ are very similar to the charged AdS black holes case so we just present the final answer. For $\kappa_T$, one finds
\bea\label{rotatingbh1}
&& t=\fft{5-6z-5z^4+10z^5-4z^6}{4z^6}\,,\nn\\
&&p_{ex}^{\pm}=\fft{(z-1)^3(14z^4-18z^3-21z^2-25z-30)}{10z^7}\,,
\eea
whilst for the susceptibility $\beta_T$
\bea\label{rotatingbh2}
&& t=\fft{6-7z-10z^4+19z^5-8z^6}{4z^5(2z-1)}\,,\nn\\
&& p_{ex}^{\pm}=\fft{(z-1)^3(28z^4-20z^3-24z^2-29z-35)}{10z^6(2z-1)}\,.
\eea
\begin{figure}
\centering
\includegraphics[width=210pt]{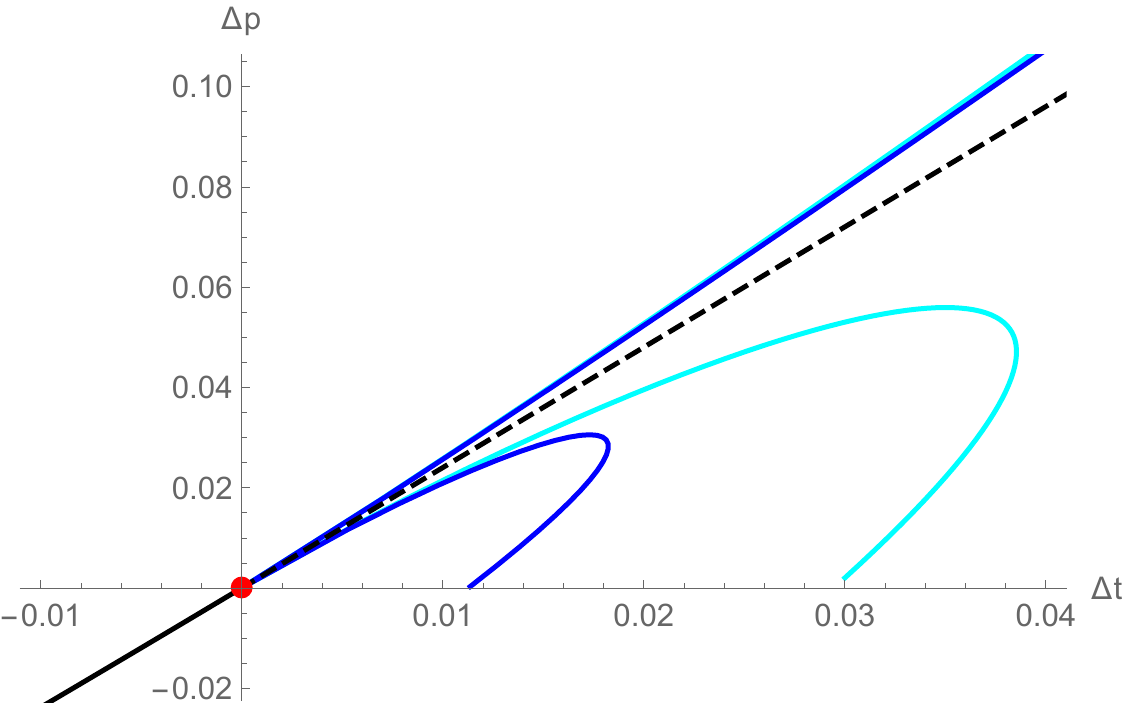}
\includegraphics[width=210pt]{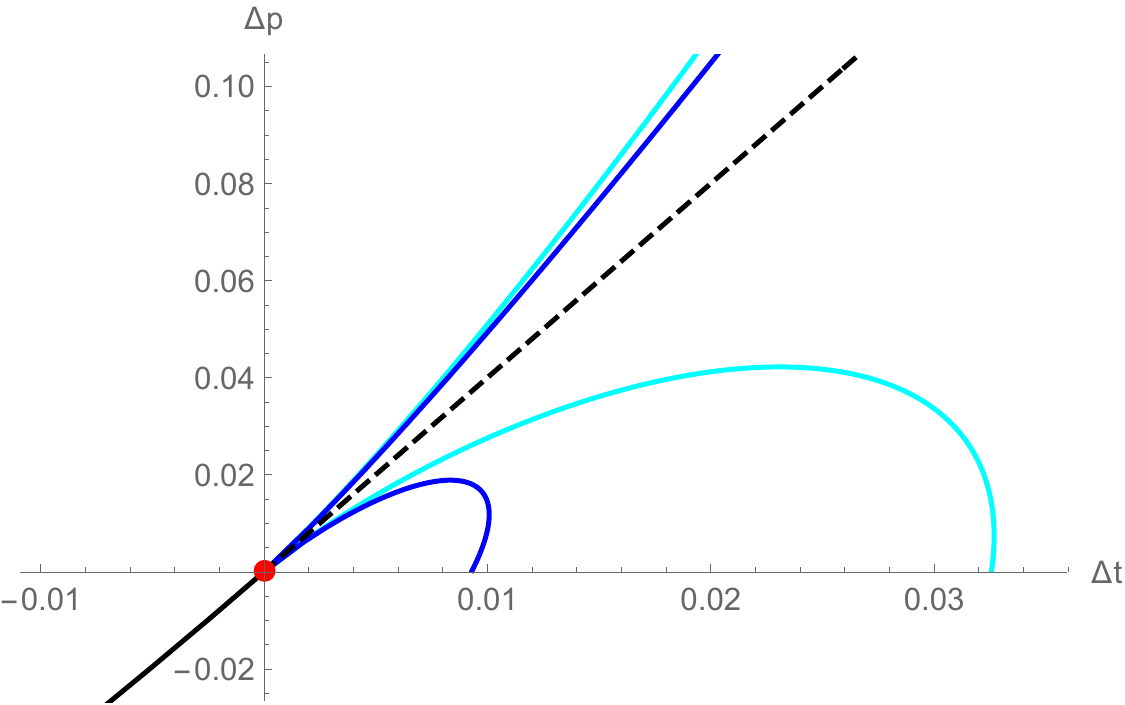}
\caption{The thermodynamic crossovers  for the rotating AdS black hole (left) and the Gauss-Bonnet black holes (right) $\Delta p=p-1\,,\Delta t=\hat t-1$. The black line is the coexistence curve. The cyan (blue) lines are the supercritical crossovers defined by using the response function $\kappa_T$ ($\beta_T$).  The dashed line stands for the critical isochore.}
\label{other}\end{figure}
The behaviors are depicted in the left panel of Fig. \ref{other}. In the critical domain, one has to leading order
\be p_{ex}^{\pm}=\pm\ft{16\sqrt{10}}{25}\, t^{3/2} \,,\quad \omega^{\pm}=\mp\ft{\sqrt{10}}{5}\,  t^{1/2}  \,.\ee

For the five dimensional Gauss-Bonnet black holes, the Locs in the extended phase space is given by \cite{Cai:2013qga}
\bea
\hat t=\fft{z(3+p z^2)}{3 z^2+1}\,.
\eea
There exists a critical point below which the coexistence curve on the $T-P$ plane reads \cite{Cui:2025bfr}
\bea
\hat{t}_*=\sqrt{\fft{p_*(3-p_*)}{2}}\,.
\eea
The thermodynamic crossovers $L^{\pm}$ are given by for $\kappa_T$
\bea\label{GB1}
&& t=-\fft{(z-1)^2(2z^2-4z-1)}{2z(z^3+3z-1)}\,\nn\\
&&p_{ex}^{\pm}=\fft{(z-1)^3(6z^4-14z^3-12z^2-3z-1)}{2z^4(z^3+3z-1)}\,,
\eea
and for $\beta_T$
\bea\label{GB2}
&&t=-\fft{(z-1)^28z^3-11z^2-2z-1}{8z^5-3z^4+16z^3-8z^2-1}\,,\nn\\
&&p_{ex}^{\pm}=\fft{3(z-1)^3(8z^4-11z^3-9z^2-3z-1)}{z^2(8z^5-3z^4+16z^3-8z^2-1)}\,.
\eea
The behaviors are depicted in the right panel of Fig. \ref{other}.
In the critical domain, one has to leading order
\be p_{ex}^{\pm}=\pm8\sqrt{2}\,t^{3/2} \,,\quad \omega^{\pm}=\mp\sqrt{2}\, t^{1/2}  \,.\ee
For these examples, the various features of the $L^\pm$ lines are again analogous to those of the VdW fluid. In particular, the $Z_2$ symmetry emerges in the critical domain, as the VdW fluid.

\newpage

\bibliographystyle{utphys}
\bibliography{reference}

\providecommand{\href}[2]{#2}\begingroup\raggedright\begin{thebibliography}{10}

\bibitem{xu2005relation}
L.~Xu, P.~Kumar, S.~V. Buldyrev, S.-H. Chen, P.~H. Poole, F.~Sciortino, and
  H.~E. Stanley, ``Relation between the widom line and the dynamic crossover in
  systems with a liquid--liquid phase transition,'' {\em Proceedings of the
  National Academy of Sciences} {\bfseries 102} no.~46, (2005) 16558--16562.

\bibitem{simeoni2010widom}
G.~Simeoni, T.~Bryk, F.~Gorelli, M.~Krisch, G.~Ruocco, M.~Santoro, and
  T.~Scopigno, ``The widom line as the crossover between liquid-like and
  gas-like behaviour in supercritical fluids,'' {\em Nature Physics} {\bfseries
  6} no.~7, (2010) 503--507.

\bibitem{ruppeiner2012thermodynamic}
G.~Ruppeiner, A.~Sahay, T.~Sarkar, and G.~Sengupta, ``Thermodynamic geometry,
  phase transitions, and the widom line,'' {\em Physical Review E} {\bfseries
  86} no.~5, (2012) 052103.

\bibitem{luo2014behavior}
J.~Luo, L.~Xu, E.~Lascaris, H.~E. Stanley, and S.~V. Buldyrev, ``Behavior of
  the widom line in critical phenomena,'' {\em Physical Review Letters}
  {\bfseries 112} no.~13, (2014) 135701.

\bibitem{corradini2015widom}
D.~Corradini, M.~Rovere, and P.~Gallo, ``The widom line and dynamical crossover
  in supercritical water: Popular water models versus experiments,'' {\em The
  Journal of Chemical Physics} {\bfseries 143} no.~11, (2015) 114502.

\bibitem{gallo2014widom}
P.~Gallo, D.~Corradini, and M.~Rovere, ``Widom line and dynamical crossovers as
  routes to understand supercritical water,'' {\em Nature communications}
  {\bfseries 5} no.~1, (2014) 1--6.

\bibitem{de2021widom}
E.~de~Jes{\'u}s, J.~Torres-Arenas, and A.~Benavides, ``Widom line of real
  substances,'' {\em Journal of Molecular Liquids} {\bfseries 322} (2021)
  114529.

\bibitem{brazhkin2012two}
V.~Brazhkin, Y.~D. Fomin, A.~Lyapin, V.~Ryzhov, and K.~Trachenko, ``Two liquid
  states of matter: A new dynamic line on a phase diagram,'' {\em Physical
  Review E} {\bfseries 85} no.~3, (2012) 031203.

\bibitem{yoon2018two}
T.~J. Yoon, M.~Y. Ha, W.~B. Lee, and Y.-W. Lee, ``“two-phase”
  thermodynamics of the frenkel line,'' {\em The Journal of Physical Chemistry
  Letters} {\bfseries 9} no.~16, (2018) 4550--4554.

\bibitem{bolmatov2014structural}
D.~Bolmatov, D.~Zav’Yalov, M.~Gao, and M.~Zhernenkov, ``Structural evolution
  of supercritical co2 across the frenkel line,'' {\em The journal of physical
  chemistry letters} {\bfseries 5} no.~16, (2014) 2785--2790.

\bibitem{bolmatov2015frenkel}
D.~Bolmatov, M.~Zhernenkov, D.~Zav’yalov, S.~N. Tkachev, A.~Cunsolo, and
  Y.~Q. Cai, ``The frenkel line: a direct experimental evidence for the new
  thermodynamic boundary,'' {\em Scientific reports} {\bfseries 5} no.~1,
  (2015) 1--10.

\bibitem{fomin2018dynamics}
Y.~D. Fomin, V.~Ryzhov, E.~Tsiok, J.~Proctor, C.~Prescher, V.~Prakapenka,
  K.~Trachenko, and V.~Brazhkin, ``Dynamics, thermodynamics and structure of
  liquids and supercritical fluids: crossover at the frenkel line,'' {\em
  Journal of Physics: Condensed Matter} {\bfseries 30} no.~13, (2018) 134003.

\bibitem{yang2015frenkel}
C.~Yang, V.~Brazhkin, M.~Dove, and K.~Trachenko, ``Frenkel line and solubility
  maximum in supercritical fluids,'' {\em Physical Review E} {\bfseries 91}
  no.~1, (2015) 012112.

\bibitem{Li_2024}
X.~Li and Y.~Jin, ``Thermodynamic crossovers in supercritical fluids,'' {\em
  Proceedings of the National Academy of Sciences} {\bfseries 121} no.~18,
  (Apr., 2024) 1091--6490.

\bibitem{kadanoff1976scaling}
L.~P. Kadanoff, ``Scaling, universality and operator algebras,'' in {\em Phase
  Transitions and Critical Phenomena}, C.~Domb and M.~S. Green, eds., vol.~5A,
  ch.~1.
\newblock Academic, New York, 1976.

\bibitem{fisher1983scaling}
M.~E. Fisher, ``Scaling, universality and renormalization group theory,'' in
  {\em Critical phenomena}, pp.~1--139.
\newblock Springer, 1983.

\bibitem{Wang:2025ctk}
S.~Wang, X.~Li, Y.~Jin, and L.~Li, ``{Analogous supercritical crossovers in
  black holes and water},'' \href{http://arxiv.org/abs/2506.10808}{{\ttfamily
  arXiv:2506.10808 [gr-qc]}}.

\bibitem{Kubiznak:2012wp}
D.~Kubiznak and R.~B. Mann, ``{P-V criticality of charged AdS black holes},''
  \href{http://dx.doi.org/10.1007/JHEP07(2012)033}{{\em JHEP} {\bfseries 07}
  (2012) 033}, \href{http://arxiv.org/abs/1205.0559}{{\ttfamily arXiv:1205.0559
  [hep-th]}}.

\bibitem{Cui:2025bfr}
H.-M. Cui and Z.-Y. Fan, ``{Analytical approach to criticality of AdS black
  holes},'' \href{http://arxiv.org/abs/2506.20959}{{\ttfamily arXiv:2506.20959
  [gr-qc]}}.

\bibitem{Cai:1998vy}
R.-G. Cai and K.-S. Soh, ``{Topological black holes in the dimensionally
  continued gravity},''
  \href{http://dx.doi.org/10.1103/PhysRevD.59.044013}{{\em Phys. Rev. D}
  {\bfseries 59} (1999) 044013},
  \href{http://arxiv.org/abs/gr-qc/9808067}{{\ttfamily arXiv:gr-qc/9808067}}.

\bibitem{Chamblin:1999tk}
A.~Chamblin, R.~Emparan, C.~V. Johnson, and R.~C. Myers, ``{Charged AdS black
  holes and catastrophic holography},''
  \href{http://dx.doi.org/10.1103/PhysRevD.60.064018}{{\em Phys. Rev. D}
  {\bfseries 60} (1999) 064018},
  \href{http://arxiv.org/abs/hep-th/9902170}{{\ttfamily arXiv:hep-th/9902170}}.

\bibitem{Gunasekaran:2012dq}
S.~Gunasekaran, R.~B. Mann, and D.~Kubiznak, ``{Extended phase space
  thermodynamics for charged and rotating black holes and Born-Infeld vacuum
  polarization},'' \href{http://dx.doi.org/10.1007/JHEP11(2012)110}{{\em JHEP}
  {\bfseries 11} (2012) 110}, \href{http://arxiv.org/abs/1208.6251}{{\ttfamily
  arXiv:1208.6251 [hep-th]}}.

\bibitem{Spallucci:2013osa}
E.~Spallucci and A.~Smailagic, ``{Maxwell's equal area law for charged
  Anti-deSitter black holes},''
  \href{http://dx.doi.org/10.1016/j.physletb.2013.05.038}{{\em Phys. Lett. B}
  {\bfseries 723} (2013) 436--441},
  \href{http://arxiv.org/abs/1305.3379}{{\ttfamily arXiv:1305.3379 [hep-th]}}.

\bibitem{Hu:2024ldp}
Y.-P. Hu, Y.-S. An, G.-Y. Sun, W.-L. You, D.-N. Shi, H.~Zhang, X.~Chen, and
  R.-G. Cai, ``{Quantum anomaly triggers the violation of scaling laws in
  gravitational system},'' \href{http://arxiv.org/abs/2410.23783}{{\ttfamily
  arXiv:2410.23783 [gr-qc]}}.

\bibitem{stanley1971}
H.~E. Stanley, {\em {Introduction to Phase Transitions and Critical
  Phenomena}}.
\newblock Oxford University Press, 1971.

\bibitem{Cui:2025qdy}
H.-M. Cui and Z.-Y. Fan, ``{Critical phenomenon of quantum BTZ black holes},''
  \href{http://arxiv.org/abs/2505.23188}{{\ttfamily arXiv:2505.23188
  [hep-th]}}.

\bibitem{Cai:2013qga}
R.-G. Cai, L.-M. Cao, L.~Li, and R.-Q. Yang, ``{P-V criticality in the extended
  phase space of Gauss-Bonnet black holes in AdS space},''
  \href{http://dx.doi.org/10.1007/JHEP09(2013)005}{{\em JHEP} {\bfseries 09}
  (2013) 005}, \href{http://arxiv.org/abs/1306.6233}{{\ttfamily arXiv:1306.6233
  [gr-qc]}}.

\end{thebibliography}\endgroup

\end{document}